\documentclass{article}
\usepackage{spconf,amsmath,graphicx,amssymb}
\usepackage[export]{adjustbox}
\usepackage{subfig}
\usepackage{color}
\usepackage{enumitem}
\usepackage{multirow}
\PassOptionsToPackage{hyphens}{url}\usepackage{hyperref}

\hypersetup{colorlinks=true}

\title{Speaker Diarization with LSTM}

\name{Quan Wang\textsuperscript{1} \quad Carlton Downey\textsuperscript{2} \quad Li Wan\textsuperscript{1} \quad Philip Andrew Mansfield\textsuperscript{1} \quad Ignacio Lopez Moreno\textsuperscript{1}\thanks{More information of this work can be found at: \url{https://google.github.io/speaker-id/publications/LstmDiarization}}}

\address{\textsuperscript{1}Google Inc., USA \qquad \textsuperscript{2}Carnegie Mellon University, USA\\[4pt] {
  \normalsize
  \textsuperscript{1}
    \{
    \href{mailto:quanw@google.com}{\nolinkurl{quanw}},
    \href{mailto:liwan@google.com}{\nolinkurl{liwan}},
    \href{mailto:memes@google.com}{\nolinkurl{memes}},
    \href{mailto:elnota@google.com}{\nolinkurl{elnota}}
    \}
  {\tt @google.com} \qquad
  \textsuperscript{2}
  \href{mailto:cmdowney@cs.cmu.edu}{\nolinkurl{cmdowney@cs.cmu.edu}}
}}

\begin{document}
\ninept
\maketitle
\begin{abstract}
For many years, i-vector based audio embedding techniques were the dominant approach
for speaker verification and speaker diarization applications.
However, mirroring the rise of deep learning in various domains, neural network based
audio embeddings, also known as \textit{d-vectors}, have consistently demonstrated superior
speaker verification performance.
In this paper, we build on the success of d-vector based speaker verification systems to develop a new d-vector based approach to speaker diarization.
Specifically, we combine LSTM-based d-vector audio embeddings with recent work in non-parametric clustering to obtain a state-of-the-art speaker diarization system.
Our system is evaluated on three standard public datasets,
suggesting that d-vector based diarization systems offer significant advantages over traditional
i-vector based systems.
We achieved a 12.0\% diarization error rate on NIST SRE 2000 CALLHOME,
while our model is trained with out-of-domain data from voice search logs.
\end{abstract}
\section{Introduction}
\label{sec:intro}

Speaker diarization is the process of partitioning an input audio stream into homogeneous segments according to the speaker identity. It answers the question ``\textit{who spoke when}" in a multi-speaker environment. It has a wide variety of applications including multimedia information retrieval,
speaker turn analysis, and audio processing. In particular, the speaker boundaries
produced by diarization systems have the potential to significantly improve acoustic speech recognition (ASR) accuracy.

A typical speaker diarization system usually consists of four components: (1) Speech segmentation, where the input audio is segmented into short sections that are assumed to have a single speaker, and the non-speech sections are filtered out; (2) Audio embedding extraction, where specific features such as MFCCs \cite{kenny2010diarization}, speaker factors \cite{castaldo2008stream}, or i-vectors \cite{shum2013unsupervised,senoussaoui2014study,sell2014speaker} are extracted from the segmented sections; (3) Clustering, where the number of speakers is determined, and the extracted audio embeddings are clustered into these speakers; and optionally (4) Resegmentation \cite{sell2015diarization}, where the clustering results are further refined to produce the final diarization results.

In recent years, neural network based audio embeddings (d-vectors) have seen  wide-spread use in speaker verification
applications \cite{variani2014deep,chen2015locally,heigold2016end,attention,ge2e},
often significantly outperforming previously state-of-the-art techniques based on i-vectors.
However, most of these applications belong to text-dependent speaker verification,
where the speaker
embeddings are extracted from specific detected keywords \cite{chen2014small,prabhavalkar2015automatic}. In contrast, speaker diarization requires text-independent embeddings which work on arbitrary speech.

In this paper, we explore a text-independent d-vector based approach to speaker diarization. We leverage the work of \cite{ge2e} to train an LSTM-based text-independent speaker verification model, then combine this model with recent work in non-parametric spectral clustering algorithm to obtain a state-of-the-art speaker diarization system.

While several authors have had explored using neural network embeddings for diarization tasks, their work has largely focused on using feed-forward DNNs to directly perform diarization. For example, \cite{garcia2017speaker} uses DNN embeddings trained on PLDA-inspired loss.
In contrast, our work uses RNNs (specifically LSTMs \cite{hochreiter1997long}),
which better capture the sequential nature of
audio signals, and our generalized end-to-end training architecture directly simulates the enroll-verify run-time logic.

There have been several attempts to apply spectral clustering \cite{von2007tutorial}
to the speaker diarization problem \cite{ning2006spectral,shum2013unsupervised}. However, to the authors' knowledge, our work is the first to combine LSTM-based d-vector embeddings with spectral clustering. Furthermore, as part of our spectral clustering algorithm, we present a novel sequence of affinity matrix refinement steps which act to de-noise the affinity matrix, and are crucial to the success of our system.

The remainder of this paper is organized as follows: In Sec. \ref{sec:dvector}, we describe how the LSTM-based text-independent speaker verification model trained with the framework in \cite{ge2e} can be adapted to featurize raw audio data and prepare it for clustering. In Sec. \ref{sec:clustering}, we describe four different clustering algorithms and discuss the pros and cons of each in the context of speaker diarization, culminating with a modified spectral clustering algorithm. Experimental results and discussions are
presented in Sec. \ref{sec:exp}, and conclusions are in Sec. \ref{sec:conclusions}.

\section{Diarization with D-Vectors}
\label{sec:dvector}
Wan \textit{et al.} recently introduced an LSTM-based \cite{hochreiter1997long}
speaker embedding network for both text-dependent and text-independent speaker verification \cite{ge2e}. Their model is trained on fixed-length segments extracted from a large corpus of arbitrary speech. They showed that the d-vector embeddings
produced by such networks usually significantly outperform i-vectors in an enrollment-verification 2-stage application. We now describe how this model can be modified for purposes of speaker diarization.

\begin{figure}
  \centering
    \includegraphics[width=0.49\textwidth]{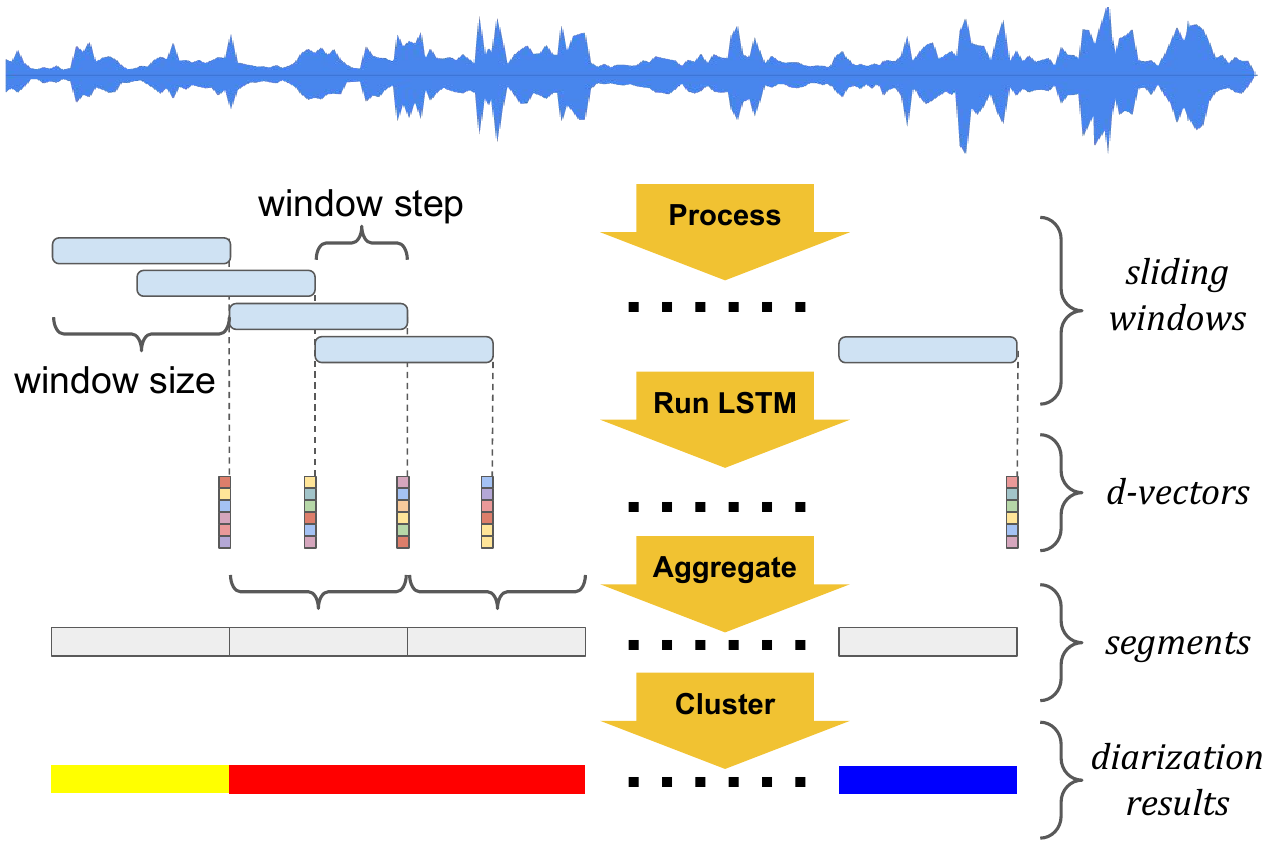}
  \caption{A flowchart of our d-vector based diarization system.}
  \label{fig:inference}
  \vspace{-10pt}
\end{figure}

The flowchart of our diarization system is provided in Fig. \ref{fig:inference}.
In this system, audio signals are first transformed into frames of width 25ms and step 10ms,
and log-mel-filterbank energies of dimension 40 are extracted from each frame as the network input.
We build sliding windows of a fixed length on these frames, and run the LSTM network
on each window.
The last-frame output of the LSTM is then used as the d-vector representation of this sliding window.

We use a Voice Activity Detector (VAD) to determine speech segments from the audio, which are further divided into smaller \textit{non-overlapping} segments using a maximal segment-length limit (\textit{e.g.} 400ms in our experiments), which determines the temporal resolution of the diarization results. For each segment, the corresponding d-vectors are first
L2 normalized, then averaged to form an embedding of the segment.

The above process serves to reduce arbitrary length audio input into a sequence of fixed-length embeddings. We can now apply a clustering algorithm to these embeddings in order to determine the number of unique speakers, and assign each part of the audio to a specific speaker.

\section{Clustering}
\label{sec:clustering}
In this section, we introduce the four clustering algorithms that we integrated into our diarization system. We place particular focus on the spectral offline clustering algorithm, which significantly outperformed the alternative approaches across experiments.

We note that clustering algorithms can be separated into two categories according to the run-time latency:
\begin{itemize}
  \item {\bf Online clustering}: A speaker label is immediately emitted once a segment is available, without seeing future segments.
  \item {\bf Offline clustering}: Speaker labels are produced after the embeddings of all segments
  are available.
\end{itemize}
Offline clustering algorithms typically outperform Online clustering algorithms due to the additional contextual information available in the offline setting.
Furthermore, a final resegmentation step can only be applied in the offline setting.
Nonetheless, the choice between online and offline depends primarily on the nature of the
application --- where the system is intended to be deployed.
For example, latency-sensitive applications such as live video analysis typically restrict the system to online clustering algorithms.

\subsection{Naive online clustering}
\label{sec:naive}

This is a prototypical online clustering algorithm. We apply a
threshold on the similarities between embeddings of segments.
To be consistent with the generalized end-to-end training architecture \cite{ge2e},
\textit{cosine similarity} is used as our similarity metric.

In this clustering algorithm, each cluster is represented
by the centroid of all its corresponding embeddings. When a new segment embedding is available,
we compute its similarities to centroids of all existing clusters. If they are all smaller than
the threshold, then create a new cluster containing only this embedding;
otherwise, add this embedding to the most similar cluster and update the centroid.

\subsection{Links online clustering}
\label{sec:links}

\textit{Links} is an online clustering method we developed to improve upon the naive approach. It estimates cluster probability distributions and models their substructure based on the embedding vectors received so far. The technical details are described in a separate paper \cite{mansfield2018links}.

\subsection{K-Means offline clustering}
\label{sec:kmeans}

Like in many diarization systems \cite{ben2012initialization,shum2013unsupervised,dimitriadis2017developing},
we integrated the K-Means clustering algorithm with our system. Specifically,
we use K-Means++ for initialization \cite{arthur2007k}.
To determine the number of speakers $\widetilde{k}$,
we use the ``elbow" of the derivatives of
conditional Mean Squared Cosine Distances\footnote{We define cosine distance as $d(x,y)=\big(1-\cos(x,y)\big)/2$.}
(MSCD) between each embedding to its
cluster centroid:
\begin{equation}
\widetilde{k}=\arg\max_{k \geq 1} | \mathrm{MSCD} ' (k) | .
\end{equation}

\subsection{Spectral offline clustering}
\label{spectral}

\begin{figure*}
  \vspace{-5pt}
  \centering
    \includegraphics[width=0.99\textwidth]{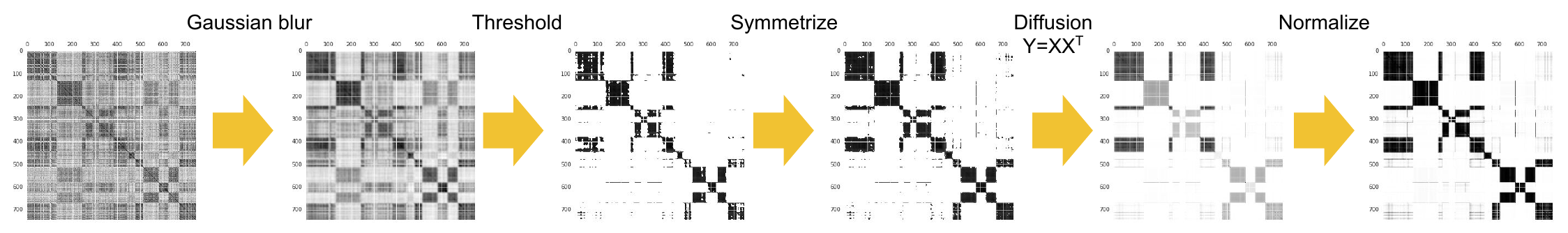}
  \caption{Refinement operations on the affinity matrix.}
  \label{fig:refine}
  \vspace{-10pt}
\end{figure*}

Our spectral clustering algorithm consists of the following steps:
\begin{enumerate}
\item Construct the affinity matrix $A$, where $A_{ij}$ is the
cosine similarity between $i$th and $j$th segment embedding when $i \neq j$, and the diaginal
elements are set to the maximal value in each row: $A_{ii}=\max_{j \neq i} A_{ij}$.

\item Apply the following sequence of refinement operations on the affinity matrix $A$:
\begin{enumerate}
  \item Gaussian Blur with standard deviation $\sigma$;
  \item Row-wise Thresholding: For each row, set elements smaller than this row's
  $p$-percentile to 0;
  \footnote{In practice, it's better to use soft thresholding: scale these elements by a small multiplier such as $0.01$.}
  \item Symmetrization: $Y_{ij}=\max(X_{ij},X_{ji})$;
  \item Diffusion: $Y=XX^T$;
  \item Row-wise Max Normalization:
  $Y_{ij}=X_{ij}/\max_{k}X_{ik}$.
\end{enumerate}
These refinements act to both smooth and denoise the data in the similarity space as shown in Fig. \ref{fig:refine}, and are crucial to the success of the algorithm. The refinements are based on the temporal locality of speech data --- contiguous speech segments should have similar embeddings, and hence similar values in the affinity matrix.

We now provide the intuition behind each of these operations: The Gaussian blur acts to smooth the data, and reduce the effect of outliers. Row-wise thresholding serves to zero-out affinities between embeddings belonging to two different speakers. Symmetrization restores matrix symmetry which is crucial to the spectral clustering algorithm. The diffusion steps draws inspiration from the Diffusion Maps algorithm \cite{coifman2006diffusion}, and serves to sharpen the image resulting in clear boundaries between sections of the affinity matrix belonging to distinct speakers. Finally, the row-wise max normalization serves to rescale the spectrum of the matrix to ensure undesirable scale effects do not occur during the subsequent spectral clustering step.

\item After all refinement operations have been applied,
perform eigen-decomposition on the refined affinity matrix.
Let the $n$ eigen-values be:
$\lambda_1 > \lambda_2 > \cdots > \lambda_n$.
We use the maximal eigen-gap to determine the number of clusters $\widetilde{k}$:
\begin{equation}
\widetilde{k}=\arg\max_{1 \leq k \leq n} \frac{\lambda_k}{\lambda_{k+1}} .
\end{equation}

\item Let the eigen-vectors corresponding to the largest $\widetilde{k}$
eigen-values be
$v_1, v_2, \cdots, v_{\widetilde{k}}$.
We replace the $i$th segment embedding by the corresponding dimension in these eigen-vectors:
$e_i=[v_{1i}, v_{2i}, \cdots, v_{\widetilde{k}i}]$.
Then we use the same K-Means algorithm in Sec. \ref{sec:kmeans} to cluster these
new embeddings, and produce speaker labels.
\end{enumerate}

\subsection{Discussion}
Speech data analysis is an extremely challenging problem domain, and conventional clustering algorithms such as K-Means often perform poorly. This is due to a number of unfortunate properties inherent to speech data, which include:

\begin{enumerate}[label=(\roman*)]
  \item \textbf{Non-Gaussian Distributions}: Speech data are often Non-Gaussion. In this setting, the centroid of a cluster (central to K-Means clustering)
        is not a sufficient representation.
  \item \textbf{Cluster Imbalance}: In speech data, it is often the case that one speaker will speak often, while other speakers will speak rarely. In this setting, K-Means may incorrectly split large clusters into several smaller clusters.
  \item \textbf{Hierarchical Structure}: Speakers fall into various groups according to gender, age, accent, \textit{etc.} This structure is problematic since the difference between a male and a female speaker is much larger than the difference between two female speakers. This makes it difficult for K-Means to distinguish between clusters corresponding to groups, and clusters corresponding to distinct speakers. In practice, this often causes K-Means to incorrectly cluster all embeddings corresponding to male speakers into one cluster, and all embeddings corresponding to female speakers into another.
\end{enumerate}

The problems caused by these properties are not limited to K-Means clustering, but are endemic to most parametric clustering algorithms. Fortunately, these problems can be mitigated by employing a non-parametric connection-based clustering algorithm such as spectral clustering.

\section{Experiments}
\label{sec:exp}

\begin{table*}
\centering
  \caption{DER (\%) on two English-only datasets for different embeddings and clustering
  algorithms.
  }
  \label{tab:der}
  \begin{tabular}{| c | c | c | c | c | c | c | c | c | c |}
    \hline
    \multirow{2}{*}{\bf Embedding} & \multirow{2}{*}{\bf Clustering}
    & \multicolumn{4}{|c|}{\bf CALLHOME American English Eval}
    & \multicolumn{4}{|c|}{\bf NIST RT-03 English CTS Eval} \\ \cline{3-10}
    & & Confusion & FA & Miss & Total & Confusion & FA & Miss & Total \\ \hline
    \multirow{4}{*}{i-vector} & Naive
    & 26.41 & \multirow{4}{*}{2.40} & \multirow{4}{*}{3.55} & 32.36 & 35.35 & \multirow{4}{*}{4.66} & \multirow{4}{*}{2.62} & 42.63 \\
    & Links
    & 25.40 & & & 31.36 & 33.56 & & & 40.48 \\
    & K-Means
    & 22.86 & & & 28.81 & 24.38 & & & 31.66 \\
    & Spectral
    & 14.59 & & & 20.54 & 13.84 & & & 21.12 \\ \hline
    \multirow{4}{*}{d-vector} & Naive
    & 12.41 & \multirow{4}{*}{1.94} & \multirow{4}{*}{4.51} & 18.87 & 18.76 & \multirow{4}{*}{4.09} & \multirow{4}{*}{4.45} & 27.30 \\
    & Links
    & 11.02 & & & 17.47 & 18.56 & & & 27.10 \\
    & K-Means
    & 7.29 & & & 13.75 & 7.80 & & & 16.34 \\
    & Spectral
    & 6.03 & & & 12.48 & 3.76 & & & 12.30 \\ \hline
  \end{tabular}
\end{table*}

\subsection{Models}
We run experiments with all combinations of both i-vector and d-vector models, with the four clustering algorithms discussed in Sec. \ref{sec:clustering}.
Both models are trained on an anonymized collection of voice searches,
which has around 36M utterances and 18K speakers.

The i-vector model is trained using 13 PLP coefficients with delta and delta-delta coefficients.
The GMM-UBM includes 512 Gaussians,
and the total variability matrix includes 100 eigen-vectors.
The final i-vectors are reduced to 50-dimensional using LDA.

The d-vector model is a 3-layer LSTM network with a final linear layer. Each LSTM layer has
768 nodes, with projection \cite{sak2014long} of 256 nodes.

Our Voice Activity Detection (VAD) model is a very small GMM model using the same PLP features
as i-vector. It only has two full covariance Gaussians:
one for speech, and one for non-speech. We found this simple VAD generalizes better
across domains (from queries to telephone) for diarization than
CLDNN \cite{zazo2016feature} VAD models.

\subsection{Datasets}
We report Diarization Error Rates (DER) on three standard public datasets:
(1) CALLHOME American English \cite{canavan1997callhome} (LDC97S42 + LDC97T14);
(2) 2003 NIST Rich Transcription (LDC2007S10),
the English conversational telephone speech (CTS) part;
and (3) 2000 NIST Speaker Recognition Evaluation (LDC2001S97), Disk-8.

The first two datasets are English only, and are relatively smaller. Thus we use these two datasets to compare different algorithms.

The third dataset is used by most diarization papers, and is usually directly referred to as ``CALLHOME" in literature. It contains 500 utterances distributed across six languages: Arabic, English, German, Japanese, Mandarin, and Spanish.

\subsection{Experiment setup}

Our diarization evaluation system is based on the pyannote.metrics library \cite{bredin2017pyannote}.

The CALLHOME American English dataset has a default 20-vs-20 utterances division for Dev-vs-Eval. For NIST RT-03 CTS,
we randomly divide the 72 utterances into 14-vs-58 Dev and Eval sets.
For each diarization system,
we tune the parameters such as Voice Activity Detector (VAD) threshold,
LSTM window size/step (Fig. \ref{fig:inference}), and clustering parameters on the Dev
set, and report the DER on the Eval set.

For NIST RT-03 CTS,
we only report DERs based on those provided un-partitioned evaluation map (UEM) files.
For the other two datasets, as is the standard convention in literature
\cite{castaldo2008stream,shum2013unsupervised,senoussaoui2014study,sell2015diarization,garcia2017speaker,zajic2017speaker},
we tolerate errors less than 250ms in locating segment boundaries. 

As is typical, for each audio file, multiple channels are merged into a single channel
\cite{shum2013unsupervised,sell2015diarization,dimitriadis2017developing},
and we do not process the parts that are before the first
annotation or after the last annotation. Additionally, as is standard in literature, we exclude
overlapped speech (multiple speakers speaking at the same time) from our evaluation.
For offline clustering algorithms, we constrain the system to produce at least 2 speakers.

\subsection{Results}

Our experimental results are shown in Table \ref{tab:der}, \ref{tab:der2} and \ref{tab:der3}.
We report the total DER together with its three components: False Alarm (FA), Miss,
and Confusion. FA and Miss are mostly from Voice Activity Detection errors, and
partly from the aggregation from frame-level i-vectors or window-level d-vectors to segments.
The FA and Miss differences between i-vector and d-vector are due to their different
window sizes/steps and aggregation logics.

In Table \ref{tab:der}, we can see that d-vector based diarization systems significantly
outperform i-vector based systems. For d-vector systems,
the optimal sliding window size and step are 240ms and 120ms, respectively.

We also observe that as expected, offline diarization produces significantly better results than online diarization. Specifically, online diarization predicts the incorrect number of speakers much more frequently than offline diarization.
This problem could potentially be mitigated by the addition of a ``burn-in" stage before entering the online mode.

In Table \ref{tab:der2}, we compare our d-vector + spectral clustering system with others' work on the same dataset.
Though our LSTM model is completely trained on out-of-domain and English-only data, we can still achieve state-of-the-art performance on this multilingual dataset. The performance could potentially be further improved by using in-domain training data and adding a final resegmentation step.

Additionally, in Table \ref{tab:der3}, we followed the same practice in \cite{zajic2017speaker} to evaluate our system
on a subset of 109 utterances from CALLHOME American English that have 2 speakers (called CH-109 in \cite{dimitriadis2017developing}).
Number of speakers is fixed to 2 for this evaluation.

\begin{table}
\centering
  \caption{DER (\%) on NIST SRE 2000 CALLHOME. Since we didn't do resegmentation, we report others' work by listing both with \& without Variational Bayesian (VB) resegmentation \cite{sell2015diarization}. Note that unlike others' work, our model is trained with out-of-domain data (English voice search vs. multilingual telephone speech).
  }
  \label{tab:der2}
  \begin{tabular}{| c | c c c c |}
    \hline
    \bf Method & \bf Confusion & \bf FA & \bf Miss & \bf Total \\ \hline
    d-vector + spectral & 12.0 & 2.2 & 4.6 & 18.8 \\
    Castaldo \textit{et al.} \cite{castaldo2008stream} & 13.7 & --- & ---  & --- \\
    Shum \textit{et al.} \cite{shum2013unsupervised} & 14.5 & --- & --- & --- \\
    Senoussaoui \textit{et al.} \cite{senoussaoui2014study} & 12.1 & --- & --- & --- \\
    Sell \textit{et al.} \cite{sell2015diarization} (+VB) & 13.7 (11.5) & --- & --- & --- \\
    Romero \textit{et al.} \cite{garcia2017speaker} (+VB) & 12.8 (9.9) & --- & --- & --- \\
    \hline
  \end{tabular}
\end{table}

\begin{table}
\centering
  \caption{DER (\%) on CALLHOME American English 2-speaker subset (CH-109).
  }
  \label{tab:der3}
  \begin{tabular}{| c | c c c c |}
    \hline
    \bf Method & \bf Confusion & \bf FA & \bf Miss & \bf Total \\ \hline
    d-vector + spectral & 5.97 & 2.51 & 4.06 & 12.54 \\
    Zaj{\'i}c \textit{et al.} \cite{zajic2017speaker} & 7.84 & --- & --- & --- \\
    \hline
  \end{tabular}
\end{table}

\subsection{Discussion}

Though we listed DER metrics from different papers in Table \ref{tab:der2} and \ref{tab:der3}, we find that it is difficult to fully align these numbers,
an unfortunately common problem in the diarization community.
This is due primarily to the large number of moving parts required for a functional diarization pipeline. For example, different teams
use different Voice Activity Detection marks (not publicly available), different training datasets, and different Dev sets for parameter tuning.

The evaluation protocols and software also differ from
paper to paper. Most teams exclude FA and Miss from their evaluations,
and directly refer to Confusion as their DER.
However, we observed that a poor VAD with high Miss usually filters out
the difficult parts in the speech, and makes the clustering problem much easier.
Some papers like \cite{dimitriadis2017developing} use the non-standard
Speaker Clustering Errors in frame percentage
as their metric, and also exclude FA and Miss from this error.
Additionally, it's unclear how overlapped speech is handled in some papers.

In our experiments, we do our best to ensure the comparisons are as fair as possible, and
avoid tuning parameters on Eval sets.

\section{Conclusions}
\label{sec:conclusions}

In this paper, we built on the success of d-vector based speaker verification systems to develop a new d-vector based approach to speaker diarization. Specifically, we combined LSTM-based d-vector audio embeddings with recent work in non-parametric clustering to obtain a state-of-the-art speaker diarization system. We conducted experiments on four clustering algorithms combined with both
i-vectors and d-vectors, and reported the performance on three standard public datasets: CALLHOME American English, NIST RT-03 English CTS, and NIST SRE 2000.
In general, we observed that
d-vector based systems achieve significantly lower DER than i-vector based systems.

\section{Acknowledgements}
\label{sec:ack}

We would like to thank Dr. Herv\'{e} Bredin for the continuous support with the
\texttt{pyannote.metrics} library. We would like to thank Dr. Gregory Sell and Prof. Pietro Laface for helping us
understand the evaluation datasets.
We would like to thank Yash Sheth and Richard Rose for the helpful discussions.

\newpage
\bibliographystyle{IEEEbib}
\bibliography{refs}

\end{document}